\documentclass[twocolumn, aps, 10pt, english, superscriptaddress, prl, 
amsmath, 
amssymb, floatfix, longbibliography
]{revtex4-1}
\usepackage[T1]{fontenc}
\usepackage[utf8]{inputenc}
\usepackage{color}
\usepackage[english]{babel}
\usepackage{graphicx}
\usepackage{cancel}
\usepackage{esint}
\usepackage{float}
\usepackage{tikz}
\usepackage{empheq}
\usepackage[export]{adjustbox}

\makeatletter

\usepackage{dcolumn}
\usepackage{bm}
\usepackage{soul}
\usepackage{amsfonts}
\usepackage{slashed}
\usepackage{enumerate}
\usepackage{mathtools}
\usepackage{physics}
\usepackage{amsthm}
\usepackage{inputenc}
\usepackage[normalem]{ulem}
\usepackage{svg}
\usepackage{comment}

\let\old@makecaption=\@makecaption
\usepackage{subcaption}
\let\@makecaption=\old@makecaption

\usepackage{epsfig}
\usepackage{dsfont}
\usepackage{xcolor}

\usepackage[unicode=true,pdfusetitle,bookmarks=true,bookmarksnumbered=false,bookmarksopen=false,breaklinks=false,pdfborder={0 0 0},pdfborderstyle={},backref=false,colorlinks=true]{hyperref}

\makeatother
\begin{document}

\title{Phase transitions in a non-Hermitian Su-Schrieffer-Heeger model\\ via Krylov spread complexity}


\author{E. Medina-Guerra}
\affiliation{Department of Condensed Matter Physics, Weizmann Institute of Science, Rehovot 7610001, Israel}

\author{I. V. Gornyi}
\affiliation{\mbox{Institute for Quantum Materials and Technologies, Karlsruhe Institute of Technology, 76131 Karlsruhe, Germany}}
\affiliation{\mbox{Institut für Theorie der Kondensierten Materie, Karlsruhe Institute of Technology, 76131 Karlsruhe, Germany}}

\author{Yuval Gefen}
\affiliation{Department of Condensed Matter Physics, Weizmann Institute of Science, Rehovot 7610001, Israel}
\date{\today}

\begin{abstract}
We investigate phase transitions in a non-Hermitian Su–Schrieffer–Heeger (SSH) model with an imaginary chemical potential via Krylov spread complexity and Krylov fidelity. The spread witnesses the $\mathcal{PT}$-transition for the non-Hermitian Bogoliubov vacuum of the SSH Hamiltonian, where the spectrum goes from purely real to complex (oscillatory dynamics to damped oscillations). In addition, it also witnesses the transition occurring in the $\mathcal{PT}$-broken phase, where the spectrum goes from complex to purely imaginary (damped oscillations to sheer decay). For a purely imaginary spectrum, the Krylov spread fidelity, which measures how the time-dependent spread reaches its stationary state value, serves as a probe of previously undetected dynamical phase transitions. 
\end{abstract}

\maketitle

\section{Introduction}
\label{sec:intro}
Krylov complexity plays an essential role in understanding quantum dynamics, as it measures the spread of an operator (or state) in a natural basis spanning the underlying subspace where the time evolution takes place. This complexity measure was initially used in the context of operators evolving under chaotic many-body Hamiltonians. It was hypothesized that the Lacnzos coefficients of chaotic quantum systems should 
be asymptotically linear in the ordered-basis numbering, implying, thus, an exponential growth of Krylov complexity in time~\cite{parker_universal_op}. These coefficients are obtained when the evolution-produced orthonormal basis is constructed from the repeated action of the generator of the dynamics and the initial state or operator, and their statistics have not only proven to be useful in probing chaotic dynamics alone but also to detect integrability, localization, and the transition from such phases to chaotic ones \cite{zhou2025,hashimoto_krylov_2023, balasubramanian2020quantum,rabinovici2022k,rabinovici2021operator,rabinovici2022krylovsupre,ballar2022krylov,bhata_spread,balasubramanian2023quantumchaosintegrabilitylate,quantifying_operator,assesing_the_saturation, alishahiha2024krylovcomplexityprobechaos,erdmenger_universal_2023,integrability_to_chaos,ganguli2024,Camargo_2024,baggioli2025krylovcomplexityorderparameter,huh2025krylovcomplexitymixedphase,jeong2025brickwalloneloopdeterminantspectral,Camargo_2024_2,Huh_2024_scrambling}.

Other, more standard complexity measures, such as, e.g.,  circuit complexity~\cite{chichi,haferkamp2022linear} and Nielsen complexity~\cite{nielsen,dowling2006geometry,nielsen2005geometric} can be seen as less ``canonical,'' as they require the introduction of penalty factors, universal gates, and tolerance bounds, among other properties. Relations between Krylov complexity and other complexity measures are expected as the latter is, in a way, a more natural measure. In fact, Krylov complexity has been shown to be an upper bound of circuit \cite{zhang2023building.6.L042001} and Nielsen \cite{craps} complexities, thus indicating that Krylov complexity is a \textit{bona fide} measure of complexity.

Applications of Krylov complexity beyond its initial scope have been found. For example, it has been used as a probe of topological phases in spin chains 
~\cite{caputa_spread_2023,caputa2}, dynamical phase transitions emerging from quantum quenches~\cite{krylov_dynamical_phase_bento,Afrasiar_2023,caputa2025localquencheskrylovperspective}, Trotter transitions~\cite{trotter}, the Zeno effect \cite{bhattacharya_spread_2024}, among many other applications \cite{xia2024complexity,beetar_complexity_2024,guerra2025correlationskrylovspreadnonhermitian,hornedal2022ultimate,Carabba2022quantumspeedlimits,gill2024speed,speed_limits,gautam_spread_2024,natural_basis,skin_lindb}. 
We remark that Krylov complexity has been heavily studied in the context of open systems governed by Lindbladian dynamics (see, e.g., Refs.~\cite{liu2022krylov, op1, op3, bhattacharya2022operator, bhattacharya2023krylov}). For a recent general review on Krylov complexity, see Ref.~\cite{nandy2024quantumdynamicskrylovspace}. The term \emph{Krylov complexity} is typically used in the Heisenberg picture, whereas \emph{Krylov spread} complexity or spread complexity is used in the Schrödinger picture. This work focuses on Krylov spread and uses these terms interchangeably.

Despite the aforementioned applications of Krylov complexity, its relation with measurement-induced entanglement phase transitions \cite{Li2018a, Skinner2019a, Chan2019a, Cao2019a, Szyniszewski2019a,  Li2019a, Bao2020a, Potter2022, Fisher2022, legal, Turkeshi_1, Turkeshi_2, poboiko} has not been fully understood: while the entanglement entropy is calculated in subregions of the system, Krylov complexity is calculated for the entire system (see, e.g., Ref.~\cite{mohsen} for a discussion on this matter). For example, it is known that in a monitored 1D Ising chain with a transverse magnetic field in the zero-click limit, there is an entanglement entropy transition from logarithmic law when the imaginary part of the spectrum is gapless to an area law when it is gapped \cite{zerba_measurement_2023}. For the same model, in Ref.~\cite{guerra2025correlationskrylovspreadnonhermitian}, it was shown that the second derivatives of the Krylov density with respect to the magnetic field and measurement rate display an algebraic divergent behavior when the imaginary part of the spectrum goes from gapless to gapped. Therefore, this seems to indicate that the derivatives of the Krylov complexity display either discontinuities or divergences when the spectrum undergoes certain changes, e.g., when it goes from gapful to gapless, when it displays non-analyticities, etc. This conjecture, if true, could be used to analyze other quantum phase transitions that are either difficult to calculate with more standard measures or that have not been discovered. 

In this spirit, systems that possess the parity and time-reversal ($\mathcal{PT}$-) symmetry \cite{bender_new,Bender_2015,meden_pt-symmetric_2023} are ideal for testing the above conjecture further, as their associated spectrum is real in the $\mathcal{PT}$-symmetric phase and can become complex, as well as purely imaginary, in the $\mathcal{PT}$-broken phase. To address this issue, we consider a $\mathcal{PT}$-symmetric non-Hermitian Su-Schrieffer-Heeger (SSH) model with a complex chemical potential. It was shown in Ref.~\cite{legal} that the entanglement entropy in this model obeys a volume law in the parameter space corresponding to the $\mathcal{PT}$-symmetric phase, as well as in a subregion where this phase is broken. Within the $\mathcal{PT}$-broken phase, a volume-to-area transition in the entanglement entropy was found when the spectrum goes from complex to purely imaginary. 

In this work, we investigate phase transitions in a non-Hermitian SSH model via Krylov spread complexity and Krylov fidelity. We demonstrate that the derivatives of the Krylov spread complexity calculated in the non-Hermitian Bogoliubov vacuum of the non-Hermitian SSH Hamiltonian distinguish the two types of phase transitions by displaying a non-analytic behavior across the parameter space. In addition, by implementing the Krylov fidelity introduced in Ref.~\cite{guerra2025correlationskrylovspreadnonhermitian}, which describes how the complexity reaches its stationary state, we are able to refine part of the $\mathcal{PT}$-broken phase that corresponds to a purely imaginary and gapped spectrum and find two additional dynamical phases. As we shall see, these two phases are mainly controlled by the slowest decay modes of the imaginary spectrum. This should be contrasted with the non-Hermitian Ising chain studied in Ref.~\cite{guerra2025correlationskrylovspreadnonhermitian}, where three dynamical phases were found when the imaginary part of the spectrum was gapped. As that model lacks $\mathcal{PT}$-symmetry, its spectrum can have gapless points and not whole gapless regions as in the SSH model we treat in this work. Thus, the real part of the spectrum also contributes to the emergence of the dynamical phase transitions. We also point out that Krylov complexity has been studied in systems having $\mathcal{PT}$-symmetry \cite{beetar_complexity_2024,bhattacharya_spread_2024}, where it was found that it distinguishes between the $\mathcal{PT}$-symmetric and broken phases. However, this was done mainly by studying the behavior of complexity over time, and hidden dynamical phases were not reported. 

This article is organized as follows. In Sec.~\ref{sec:model}, we define the non-Hermitian SSH Hamiltonian and describe its spectrum. In Sec.~\ref{sec:spread_complexity}, we briefly introduce the Krylov spread complexity of states. In Sec.~\ref{sec:krylov_spread_unitary_dynamics}, we discuss the Krylov spread complexity for the Hermitian evolution that transforms a particular initial state to the non-Hermitian vacuum of the SSH Hamiltonian. We demonstrate that the second derivative of the spread with respect to one of the parameters either diverges or is discontinuous at the boundaries between regions where the spectrum of the Hamiltonian changes when it goes from real to complex or purely imaginary and vice versa. In Sec.~\ref{sec:time_dependent_krylov}, we find the Krylov spread of the evolution of the same initial state used in the unitary evolution, but this time via the non-Hermitian SSH Hamiltonian when its spectrum is purely imaginary. We define the time to achieve the stationary state and find two dynamical phase transitions dictated by it. Finally, we present our conclusions and summary in Sec.~\ref{sec:summary}. 
Technical details of calculations are relegated to Appendices.

\section{Model}\label{sec:model}
We consider the SSH model with $L$ sites and with an imaginary chemical potential \cite{legal,topo_ssh}, which is described by the Hamiltonian
\begin{align}
    H&= \sum_{n=1}^{L}\left(t_1c^\dagger_{A,n}d_{B,n}+t_2c^\dagger_{A,n}d_{B,n-1}+ \text{H.c.} \right)\notag 
    \\ &\qquad \qquad +i\gamma\sum_{n=1}^L \left(c_{A,n}^\dagger c_{A,n} + d^\dagger_{B,n}d_{B,n}\right)
    \label{eq:1}
\end{align}
and obeys the anti-periodic boundary conditions.  
Here, $\gamma \in \mathbb{R}$ and $c_{A,n}$ and $d_{B,n}$ are the fermionic operators in sublattices $A$ and $B$, respectively. For convenience, we set the intracell and intercell hoppings as $t_1 = -J - h/2$ and $t_2 = -J+h/2$, respectively, where $h \in \mathbb{R}$ and $J>0$ is fixed. The anti-Hermitian part of the Hamiltonian can be thought of as the result of the backaction of a continuous measurement of the local density of particles and holes on sublattices $A$ and $B$, respectively, where no click was detected \cite{breuer2002theory}. 

The Hamiltonian \eqref{eq:1} can be diagonalized by implementing the Fourier representation (see Appendix~\ref{sec:diag})
\begin{equation}\label{eq:2}
    \begin{pmatrix}
        c_{A,n}\\
        d_{B,n}
    \end{pmatrix} = \dfrac{e^{-i\frac{\pi}{4}\sigma_x}}{\sqrt{L}}\sum_{k\in \mathcal{K}}e^{ikn}\begin{pmatrix}
         c_k \\
        d_k
    \end{pmatrix},
\end{equation}
where $\sigma_x$ is a Pauli matrix and 
\begin{equation}\label{eq:3}
     \mathcal{K} = \left\lbrace k_n = \dfrac{2\pi}{L}\left(- \Bigg\lceil \frac{L-1}{2}\Bigg\rceil + n +\frac{1}{2}\right): n \in [0,L)\cap\mathbb{N} \right\rbrace
\end{equation}
is the momenta set in the first Brillouin zone, yielding
\begin{equation}\label{eq:4}
     H = \sum_{k}\Lambda(k)\left(\chi_{+,k}^*\chi_{+,k}- \chi_{-,k}^*\chi_{-,k}\right).
\end{equation}
Here, $\chi_{\pm, k}$ are non-Hermitian quasiparticle fermionic operators obeying the anti-commutation relations 
    \begin{align}
    \{\chi_{s,k}^*, \chi_{s',k'} \} &= \delta_{ss'}\delta_{kk'}, \quad \{ \chi_{s,k}^*, \chi_{s',k'}^* \}=0, \notag \\
    \{ \chi_{s,k}, \chi_{s',k'}\}&=0. \label{eq:5}
    \end{align}
The non-Hermiticity also implies $\chi_{s,k}^\dagger \neq \chi_{s,k}^*$.

\begin{figure*}
\includegraphics[width=2\columnwidth,center]{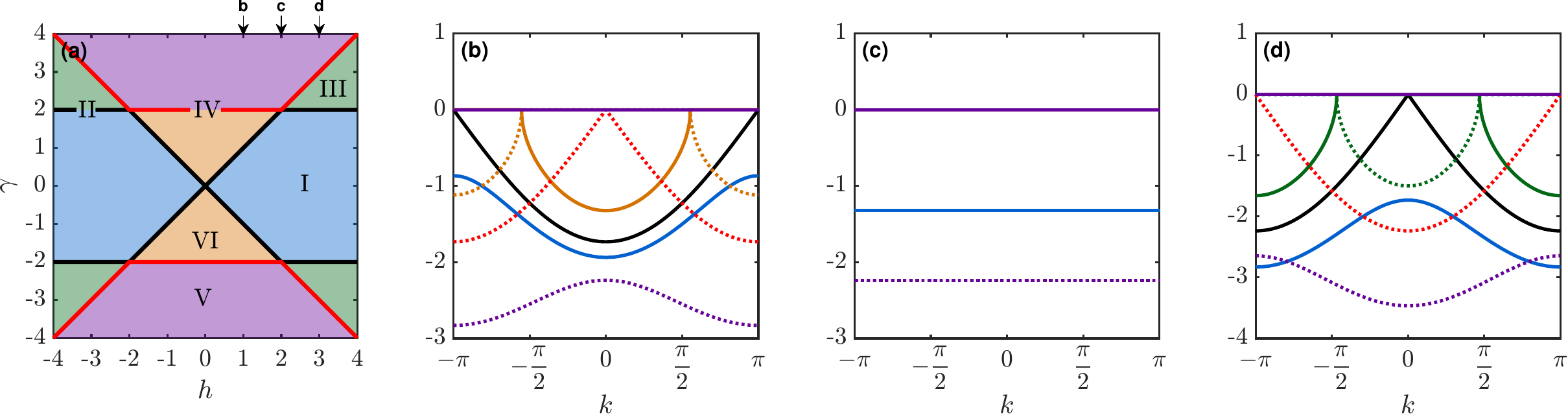}
    \caption{Negative branch $-\Lambda(k)$ of the spectrum \eqref{eq:7} in terms of the parameters $(h,\gamma)$ for $J=1$. Panel (a) displays the $h$-$\gamma$ plane with the six subsets I-VI identified in the main text. Note that regions II and IV are, respectively, all the black and all the red lines separating the two-dimensional regions (I, III, V, and VI). Panels (b)-(d) display the real (solid lines) and imaginary (dotted lines) parts of $-\Lambda(k)$ for several values of $(h,\gamma)$. Specifically, in Panel (b), $h=1$
    and $\gamma \in \{1/2,1, 3/2, 2,3\}$; in panel (c), $h=2$ and $\gamma \in \{3/2, 2, 3\}$; in panel (d), $h=3$ and $\gamma \in \{1,2,5/2, 3,4 \}$. For each of the panels (b), (c), and (d), the values of $\gamma$ go from bottom to top in the vertical cross-sections of diagram (a) indicated by arrows b, c, and d, referring to the corresponding panels. The line colors in (b)-(d) correspond to the position of the ordered pair $(h,\gamma)$ in the diagram (a), where each pair is located in one of the distinct colored regions. }
    \label{fig:1}
\end{figure*}
 The relations between the non-Hermitian operators and the sublattice operators $c$ and $d$ are given in Eqs.~\eqref{eq:ap4} and \eqref{eq:ap5}. The spectrum of $H$ is  
\begin{subequations}\label{eq:7}
\begin{align}
    \pm \Lambda(k) &= \pm \sqrt{h^2-\gamma^2+ (4J^2-h^2)\cos^2\dfrac{k}{2}}\\
    &\equiv \pm E(k) \pm i \Gamma(k),
\end{align}
\end{subequations}
 where $E(k) \coloneqq \mathfrak{Re}\, \Lambda(k)$ and $\Gamma(k) \coloneqq \mathfrak{Im}\, \Lambda(k).$ For a given $k\in \mathcal{K}$, $\Lambda(k)$ is either real or imaginary, and it completely vanishes at the points
\begin{equation}\label{eq:8}
   \pm k_{\text{EP}} = \pm 2\arccos \sqrt{\dfrac{\gamma^2-h^2}{4J^2-h^2}},
\end{equation}
which correspond to two exceptional points (EPs) \cite{Heiss_2012}.

Based on the above features, we identify the following six subregions in the $h$-$\gamma$ plane displayed in Fig.~\ref{fig:1}(a):
\begin{enumerate}[I.]
    \item $\{ \abs{\gamma} < \abs{h}, \abs{h}\leq 2J\}\cup \{\abs{\gamma} < 2J, \abs{h}\geq 2J\}$. $E(k)$ is gapped, i.e., $E(k)\neq 0$,  and $\Gamma(k) \equiv 0$.  
    \item $\{\abs{\gamma} = \abs{h}, \abs{h} \leq 2J\}\cup \{ \abs{\gamma} = 2J,  \abs{h}\geq 2J \}$. $\Gamma(k)\equiv 0$ and $E(k)$ is gapless at $\pm k_{\text{EP}}  = \pm \pi$ for $h<2J$, and at $k_{\text{EP}}  = 0$ for $h>2J$. $\Lambda(k)\equiv 0 $ for $h=\gamma=2J.$ 
    \item $\{2J<\abs{\gamma} < \abs{h}, \abs{h} > 2J \}$. $E(k)=0$ for $\abs{k} \leq k_{\text{EP}} $ and $\Gamma(k) = 0$ for $\abs{k} \geq k_{\text{EP}} $.
    \item $\{\abs{\gamma} = 2J, \abs{h}\leq 2J \}\cup \{\abs{\gamma} = \abs{h}, \abs{h}\geq 2J\}$. $E(k)\equiv 0$ and $\Gamma(k)$ is gapless at $k_{\text{EP}}  = 0$ for $h\leq 2J$, and at $\pm k_{\text{EP}}  = \pm \pi$.
    \item $\{ \abs{\gamma} >2J, \abs{h}\leq 2J \}\cup \{\abs{\gamma} >\abs{h}, \abs{h}\geq 2J \}$. $\Gamma(k)$ is gapped and $E(k)\equiv 0.$
    \item $\{ \abs{h} <\abs{\gamma} < 2J, 0<\abs{h}< 2J\}$. $\Gamma(k) = 0 $ for $\abs{k} \leq k_{\text{EP}} $ and $E(k) = 0$ for $\abs{k}  \geq k_{\text{EP}}$. 
\end{enumerate}
The reason the spectrum is purely real in regions I and II, even though the Hamiltonian \eqref{eq:1} is not Hermitian, is the $\mathcal{PT}$-symmetry \cite{legal,Rottoli_2024,bender_new,Bender_2015}, where $\mathcal{P}$ is the parity operator defined as
\begin{equation}\label{eq:9}
    \mathcal{P}c_{A,n}\mathcal{P} = d_{B,L-n+1}, \quad  \mathcal{P}d_{B,n}\mathcal{P} = c_{A,L-n+1},
\end{equation}
and $\mathcal T$ is the anti-unitary time-reversal operator defined as $\mathcal{T}\lambda \mathcal{T} = \lambda^*$, for $\lambda \in \mathbb{C}$. Thus, regions I and II correspond to a $\mathcal{PT}$-symmetric phase, and the other regions correspond to $\mathcal{PT}$-broken phases.

\section{Spread complexity of states}\label{sec:spread_complexity}
Here, we briefly review the Krylov spread complexity of states to establish further notation. We shall not address the Krylov complexity of density matrices and operators. For a recent review, see Ref.~\cite{nandy2024quantumdynamicskrylovspace}.

Let $H$ be the Hamiltonian generating the dynamics $e^{-iHt}\ket{\psi(0)}$, where $\ket{\psi(0)} \in \mathcal{H}$ is some initial state. The Krylov space, generated by the initial state and the Hamiltonian, is 
\begin{equation}
    \mathfrak{K}\big(\psi(0), H\big)\!= \text{span}\left\lbrace \ket{\psi(0)}\!, H\ket{\psi(0)}\!, \ldots , H^n\ket{\psi(0)}\!,\ldots\right\rbrace.
\end{equation}
By implementing the Gram-Schmidt orthonormalization procedure on the linearly independent set of vectors from the ordered set $\{H^n\ket{\psi(0)}\}_{n\geq 0}$, one can find the \emph{ordered} basis, called the Krylov basis, $\{ \ket{K_n}\}_n$, where $n =0,\ldots, \text{dim}\, \mathfrak{K} \leq \text{dim}\, \mathcal{H}$ and $\ket{K_0} = \ket{\psi(0)}$. Hence, the non-Hermitian and evolution can be expanded in terms of this basis as $\ket{\psi(t)} = \sum_{n=0}^{\text{dim}\,\mathfrak{K}}\varphi_n(t)\ket{K_n}$, where $\varphi_n(t) = \bra{K_n}\ket{\psi(t)}$. The Krylov spread complexity of the evolved state $\ket{\psi(t)}$ is defined as 
\begin{equation}\label{eq:9.3}
    C(\psi(0),H;t) \equiv \mathcal{C}(t) = \sum_{n=0}^{\text{dim}\,\mathfrak{K}}n\abs{\varphi_n(t)}^2,
\end{equation}
and it measures the mean value of the evolved state on the Krylov basis. As was shown in Ref.~\cite{quantum_chaos_and_complexity}, Eq.~\eqref{eq:9.3} is the spread that minimizes $\sum_n n \abs{\bra{\phi(t)}\ket{B_n}}^2$ on all possible ordered bases $\{\ket{B}_n \}_n$. 

The above construction also holds for non-Hermitian Hamiltonians and is what we implement in this work. Other approaches such as, for example, the bi-Lanczos algorithm \cite{nandy2024quantumdynamicskrylovspace,bhattacharya2023krylov} and the singular-value decomposition \cite{nandy2024,baggioli2025singularvaluedecompositionblind}, among others, focus on different ways of obtaining the Krylov basis. Those methods are implemented primarily for computational reasons, and we do not use them in our analysis. 

\section{Krylov spread via unitary dynamics}\label{sec:krylov_spread_unitary_dynamics}

In what follows, we calculate the Krylov spread density for a state that evolves unitarily to the non-Hermitian vacuum of $H$. We demonstrate that the second derivatives of the spread density with respect to $\gamma$ display a non-analytic behavior in regions II and IV, implying that this complexity measure can detect the $\mathcal{PT}$-symmetry breaking, as well as the spectral transition from a complex to a purely imaginary spectrum. The latter provides the mechanism responsible for the volume-to-area entanglement phase transition reported in Ref.~\cite{legal}. 

Consider the two Hermitian Hamiltonians 
\begin{equation}\label{eq:11}
    H_0 = -\sum_{k\in \mathcal{K}}\left( J + \frac{h}{2} \right)\sin k\left(c_k^\dagger c_k - d_k^\dagger d_k \right).
\end{equation}
and 
\begin{equation}\label{eq:13}
    H_\Omega = \sum_{k\in \mathcal{K}^+}\left[i y_+(k)c_k^\dagger d_k + i y_-(k)d^\dagger_{-k}c_{-k} + \text{H.c.} \right],
\end{equation}
where 
\begin{equation}\label{eq:14}
    y_\pm (k) = -\dfrac{t_1+ t_2\cos k  \mp  \gamma}{\sqrt{t_1^2+t_2^2-\gamma^2 +2t_1t_2\cos k}+t_2\sin k }.
\end{equation}
Let $\ket{0}$ be the vacuum annihilated by $c_k$ and $d_k$ for any $k \in \mathcal{K}$, and consider the ground state of $H_0$, 
\begin{equation}\label{eq:10}
    \ket{\psi(0)} = \ket{\text{GS}} = \prod_{k \in \mathcal{K}^+}c_{-k}^\dagger d_k^\dagger \ket{0}.
\end{equation}
It turns out that if we evolve this state via $H_\Omega$ during unit time, we reach the ground state of Eq.~\eqref{eq:1},
\begin{equation}\label{eq:12}
\ket{\Omega} = \prod_{k\in \mathcal{K}^+}\dfrac{\chi_{-,-k}^*\chi_{-,k}^*}{\bra{0} ( \chi_{-,-k}^*\chi_{-,k}^*)^\dagger \chi_{-,-k}^*\chi_{-,k}^*\ket{0}^{1/2}}\ket{0}.
\end{equation}  
As we show in Appendix~\ref{sec:diag}, $H_\Omega$ is induced by the unitary operator $\Omega \in \text{SU}(2)^{\otimes L}/\text{U}(1)^{\otimes L}$, satisfying $\ket{\Omega}  =\Omega \ket{\text{GS}}$. In other words, $\ket{\Omega}$ is a generalized coherent state of $\ket{\text{GS}}$, and $\ket{\psi(1)} = e^{-iH_\Omega}\ket{\text{GS}} = \Omega\ket{\text{GS}}$.

Moreover, as $\ket{\text{GS}}$ is the lowest-weight $\mathfrak{su}(2)\times \mathfrak{su}(2)$-state, the spread per site (or spread density) of $\ket{\psi(1)}= \ket{\Omega}$ is given by the thermodynamic limit of the sum over complexity spreads of each mode \cite{caputa2,caputa_spread_2023}:
\begin{subequations}
\begin{align}
    \mathcal{C}_\Omega &\coloneqq \lim_{L\to \infty}\dfrac{1}{L}\sum_{\substack{k\in \mathcal{K}^+\\ s = \pm }}C_s(t=1;k) \label{eq:14.1}\\
    &= \sum_{s=\pm}\int_{0}^\pi \dfrac{\dd k}{2\pi}C_s(t=1;k),\label{eq:15}
\end{align}
\end{subequations}
where $$C_s(t=1;k) = \frac{\abs{y_s(k)}^2}{1+\abs{y_s(k)}^2}$$ is the complexity per mode.
Upon replacing $y_{\pm}(k)$ and performing the integral, we get the following results for the spread in regions I-VI:
\begin{widetext}
    \begin{equation}\label{eq:15.1}
        \mathcal{C}_\Omega = \begin{cases}
            \dfrac{1}{2} - \dfrac{1}{\pi(h + 2J)}\left( \sqrt{h^2-\gamma^2}-\sqrt{4J^2-\gamma^2}-\gamma \arccot\dfrac{\gamma}{\sqrt{h^2-\gamma^2}} + \gamma \arccot \dfrac{\gamma}{\sqrt{4J^2-\gamma^2}}   \right), & \text{I+II},\\[3ex]
            \dfrac{1}{2}-\dfrac{1}{\pi(h + 2J)}\left(\sqrt{h^2-\gamma^2} - \gamma \arccot \dfrac{\gamma}{\sqrt{h^2-\gamma^2}} \right), & \text{III},  \\[3ex]
            \dfrac{1}{2},& \text{IV+V},\\[3ex]
            \dfrac{1}{2} + \dfrac{1}{\pi(h + 2J)}\left(\sqrt{4J^2-\gamma^2} - \gamma \arccot\dfrac{\gamma}{\sqrt{4J^2-\gamma^2}} \right). &  \text{VI}. 
        \end{cases}
    \end{equation}
\end{widetext}

\begin{figure*}[htb]
        \centering
        \begin{subfigure}[t]{0.44\textwidth}
            \centering
            \includegraphics[width=\linewidth, keepaspectratio]{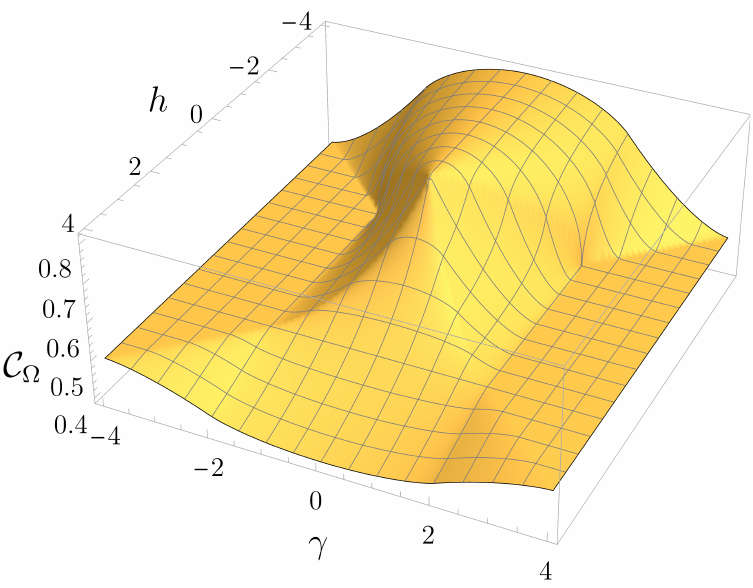}
            \caption{}
            \label{fig:2}
        \end{subfigure}
        \hfill
        \begin{subfigure}[t]{0.45\textwidth}
            \centering
            \includegraphics[width=\linewidth, keepaspectratio]{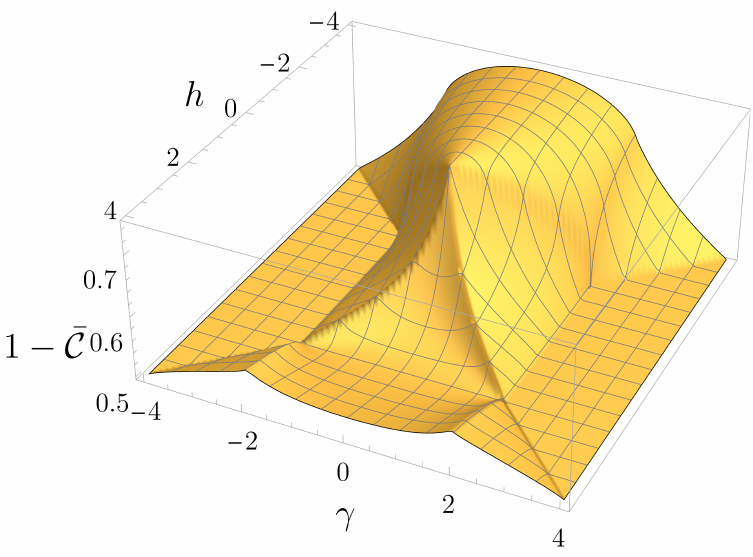}
            \caption{}
            \label{fig:2.1}
        \end{subfigure}
        \caption{(a) Krylov spread complexity density $\mathcal{C}_\Omega$ [Eq.~\eqref{eq:15}] of the unitary evolution taking $\ket{\text{GS}}$ [Eq.~\eqref{eq:10}] to  $\ket{\Omega}$ [Eq.~\eqref{eq:12}] via the Hamiltonian $H_\Omega$, Eq.~\eqref{eq:13}, at $J= 1$.  (b) $1-\bar{\mathcal{C}}$, where  $\bar{\mathcal{C}}$ is the time-averaged Krylov spread complexity density \eqref{eq:av1} of a non-Hermitian evolution of $\ket{\text{GS}}$ induced by Hamiltonian \eqref{eq:1} of the non-Hermitian SSH model. In regions IV and V (see Fig.~\ref{fig:1}), the spectrum is purely imaginary, and it can have maximally two gapless points in region IV. Thus, the state $\ket{\text{GS}}$ reaches $\ket{\Omega}$ in the infinite-time limit in the non-Hermitian evolution, and the two spreads coincide in such regions, i.e., $\mathcal{C}_\Omega = \bar{\mathcal{C}} = 1/2$. }
        \label{fig:spreads}
    \end{figure*}

\begin{figure*}
\includegraphics[width=2\columnwidth,center]{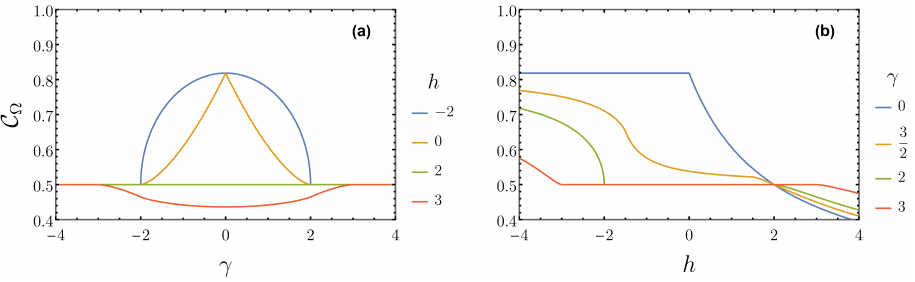}
    \caption{Transverse cross-sections of the spread $\mathcal{C}_\Omega$, Eq.~\eqref{eq:15.1}. (a) Spread as a function of $\gamma$ for $h\in \{-2,0,2,3 \}$. (b) Spread as a function of $h$ for $\gamma \in \{0,3/2,2,3\}$.  In all cases, $J=1$.}
    \label{fig:4}
\end{figure*}

There is no divergence of $\mathcal{C}_\Omega$ in regions I and II for $h = -2J$, as  
$$\mathcal{C}_\Omega\vert_{h=-2J} = \frac{1}{2} + \dfrac{\sqrt{4J^2-\gamma^2}}{2\pi J}.$$
This can be shown either by taking the limit $h\rightarrow -2J$ in Eq.~\eqref{eq:15.1} or by setting $h = -2J$ in Eq.~\eqref{eq:15}. For regions III and VI, the value of $\mathcal{C}_\Omega$ tends to $1/2$ in the limit $h\to -2J$,
where $|\gamma|\to 2J$ in these regions (see Fig,~\ref{fig:1}a). 
As we can see in Figs.~\ref{fig:2} and \ref{fig:4}, the spread is continuous across the entire domain of regions I-VI (Fig,~\ref{fig:1}a)---this appears to be a characteristic of the spread for this type of non-Hermitian systems, see, e.g., Ref.~\cite{guerra2025correlationskrylovspreadnonhermitian}.  

However, the derivatives of $\mathcal{C}_\Omega$ with respect to $h$ or $\gamma$ display singular behavior. If we start in Region I (i.e., the $\mathcal{PT}$-symmetric phase) with $h = -2J$ (blue curve in Fig.~\ref{fig:4}a), the first derivative with respect to $\gamma$ is
\begin{equation}\label{eq:di1.0}
    \left(\pdv{\mathcal{C}_\Omega\vert_{h=-2J}}{\gamma} \right)_{\text{I}} = -\dfrac{\gamma}{2J\pi\sqrt{4J^2-\gamma^2}},
\end{equation}
and it diverges at $\abs{\gamma} = 2J$. For any other point of Region I with $h\neq -2J$, the second derivative with respect to $\gamma$ is
\begin{equation}\label{eq:di1}
      \left(\pdv[2]{\mathcal{C}_\Omega}{\gamma}\right)_{\text{I}} = \dfrac{1}{\pi(h+2J)}\left( \dfrac{1}{\sqrt{4J^2-\gamma^2}}-\frac{1}{\sqrt{h^2-\gamma^2}}\right),
\end{equation}
and it diverges as Region II is approached. Furthermore, starting with $\gamma = 0$ (blue curve in Fig.~\ref{fig:4}b), which corresponds to the Hermitian case, the spread is
\begin{equation}\label{eq:d1.1}
    \left(\mathcal{C}_\Omega\vert_{\gamma = 0} \right)_{\text{I}} = \dfrac{1}{2} - \dfrac{\abs{h}-2J}{\pi(h+2J)},
\end{equation}
and it is equal to $1/2 + 1/\pi$ for $h<0$ \footnote{Note that Eq.~(16) of Ref.~\cite{caputa2} yields a value of $1/2-1/\pi$ in the same phase. This difference is due to the choice of the sign in the matrix exponential in Eq.~\eqref{eq:2}; that is, if $\sigma_x \rightarrow - \sigma_x$, we would recover that result. This sign convention determines the ``gauge'' of the non-Hermitian vacuum state $\ket{\Omega}$ and, hence, the form of $H_\Omega$, leading to different spreads from $\ket{\text{GS}}$ via the unitary evolution. This freedom, however, does not affect the presence of singularities in the derivatives of the spread.}. This phase corresponds to a non-trivial topological phase of the Hermitian SSH model, where the circle $(R_x-t_1)^2+R_z^2= t_2^2$ [see Eq.~\eqref{eq:ap1.1}] encloses the origin. Thus, the topological index is non-zero as $k$ varies in $[-\pi,\pi)$ \cite{araujo}.

Starting in Region III, the second derivative with respect to $\gamma$ is 
\begin{equation}\label{eq:di2}
    \left(\pdv[2]{\mathcal{C}_\Omega}{\gamma}\right)_{\text{III}} = -\dfrac{1}{\pi(h+2J)}\dfrac{1}{\sqrt{h^2-\gamma^2}}.
    \end{equation}
Note that there is no divergence as we approach Region II, yet there is a divergent behavior as we approach Region V. Clearly, if we start in Region V, $(C_\Omega)_{\text{V}} = 1/2$, and there is no divergence as we approach Region IV. Finally, if we start in Region VI, we have
\begin{equation}
    \left(\pdv{\mathcal{C}_\Omega}{\gamma}\right)_{\text{VI}} = -\dfrac{1}{\pi(h+2J)}\arccot\dfrac{\gamma}{\sqrt{4J^2-\gamma^2}}.
\end{equation}
For $h=0$, this derivative is discontinuous at $\gamma = 0$. The second derivative gives 
\begin{equation}\label{eq:di3}
    \left(\pdv[2]{\mathcal{C}_\Omega}{\gamma}\right)_{\text{VI}} = \dfrac{1}{\pi(h+2J)}\dfrac{1}{\sqrt{4J^2-\gamma^2}},
\end{equation}
and it starts diverging only when we approach Region IV.  

We point out that the derivatives of $\mathcal{C}_\Omega$ with respect to $h$ also display discontinuities or divergences across the phase diagram at the border lines separating distinct phases. This can be seen directly from $\mathcal{C}_\Omega$ displayed in Figs~\ref{fig:2} and~\ref{fig:4}. Thus, we conclude that the Krylov spread induced by the unitary evolution from $\ket{\text{GS}}$ to the vacuum of the non-Hermitian Hamiltonian captures, through its derivatives, the known phase transitions of the non-Hermitian SSH model.

\section{Time-dependent Krylov spread and dynamical phase transitions}\label{sec:time_dependent_krylov}

In this section, we calculate the Krylov spread of the evolution of $\ket{\psi(0)}=\ket{\text{GS}}$ via the non-Hermitian Hamiltonian $H$ in the $\mathcal{PT}$-broken phase where its spectrum is purely imaginary and gapped, i.e., $\Lambda(k) = i\Gamma(k) \neq 0\, \forall k\in \mathcal{K}$, and find $\mathcal{C}(t) \rightarrow (\mathcal{C}_\Omega)_V = 1/2$ as $t\rightarrow \infty$. We also implement the Krylov fidelity introduced in Ref.~\cite{guerra2025correlationskrylovspreadnonhermitian} to identify dynamical phases based on how the time-dependent Krylov spread reaches its infinite-time limit. We close the section by providing some comments about the relation between the above spread and the one (induced by the unitary evolution with $H_\Omega$) addressed in the previous section.  

Under the non-Hermitian Hamiltonian \eqref{eq:1}, the initial state $\ket{\psi(0)}=\ket{\text{GS}}$ evolves as follows (see Appendix~\ref{sec:time_evolution_GS}):
\begin{subequations}\label{eq:b2}
\begin{align}
    \ket{\psi(t\rightarrow \infty)} &= \lim_{t\to \infty}\dfrac{e^{-iHt}\ket{\text{GS}}}{\norm{e^{-iHt}\ket{\text{GS}}}}\\
    &= \prod_{\substack{k>0\\ s =\pm }}\dfrac{\exp[x_s(k)J_+^s(k)]}{\sqrt{1+\abs{x_s(k)}^2}}\ket{\text{GS}}, \label{eq:b2.1}
\end{align}
\end{subequations}
where 
\begin{equation}\label{eq:b3}
    x_{\pm}(k)\coloneqq \dfrac{R_x \mp i R_y}{R_z - i \Gamma(k)}.
\end{equation}
The resulting state $\ket{\psi(t\rightarrow \infty)}$ is also a coherent state of $\ket{\text{GS}}$ and is related to $\ket{\Omega}$ [Eq.~\eqref{eq:12}] via the unitary operator
\begin{equation}\label{eq:b4}
O_z\coloneqq \prod_{\substack{k>0 \\ s= \pm}}\exp[2i\phi_kJ_z^s(k)+2\phi_k]
\end{equation}
belonging to the isotropy group of $\ket{\text{GS}}$, $\text{U}(1)^{\otimes L}$, where $\phi_k = \text{Arg}\,[R_z+i\Gamma(k)]$. Specifically, $$\ket{\psi(t\rightarrow\infty)}  = O_z\ket{\Omega}= \overline{\Omega}\ket{\text{GS}} = \ket*{\overline{\Omega}},$$ where $\overline{\Omega} = O_z\Omega O_z^\dagger \in \text{SU}(2)^{\otimes L}/\text{U}(1)^{\otimes L}$. 

Since both $\ket{\Omega}$ and $\ket{\overline{\Omega}}$ are generalized coherent states of $\ket{\text{GS}}$, we can assert that time-dependent spread density of Eq.~\eqref{eq:b2} before taking the limit is
\begin{equation}\label{eq:b5}
    \mathcal{C}(t) = \sum_{s =\pm }\int_{0}^\pi \dfrac{\dd k}{2\pi}\,\dfrac{\abs{A_+^s(k;t)}^2}{1+\abs{A_+^s(k;t)}^2},
\end{equation}
and it tends to $1/2$ as $t\rightarrow \infty$ when $\Lambda(k) = i\Gamma(k)$. Note that this is precisely $\left(\mathcal{C}_\Omega \right)_{\text{V}} = 1/2$, so that the spread of $\ket{\psi(t)}$, as it evolves from $\ket{\text{GS}}$ to $\ket{\overline{\Omega}}$ via $H$ with a purely imaginary spectrum, is the same as the spread of the same initial state evolving to $\ket{\Omega}$ via the Hermitian Hamiltonian \eqref{eq:13}.

\begin{figure}[t]
\includegraphics[width=0.5\columnwidth,center]{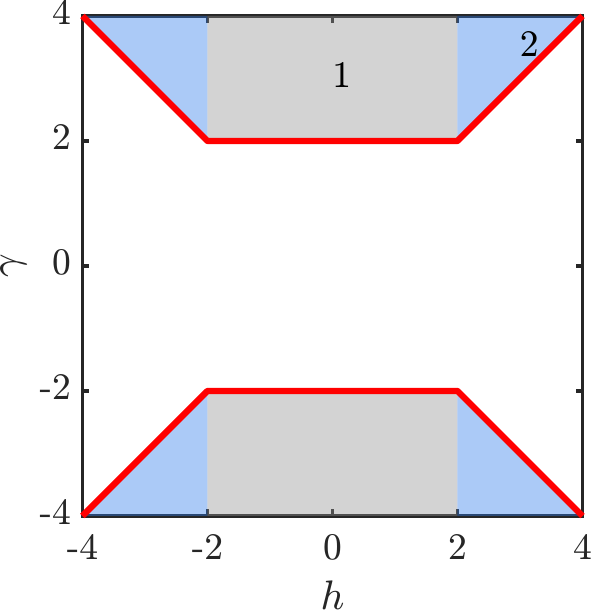}
    \caption{Schematic phase diagram defined via the Krylov fidelity \eqref{eq:b5.1} with $J=1$. Region 1 (in gray) corresponds to the time \eqref{eq:b7.1}, and Region 2 (in light blue) corresponds to the time \eqref{eq:b7.2}. 
    The points in red correspond to Region IV (see Fig.~\ref{fig:1}), where $\Lambda(k) = i\Gamma(k) = 0$, so we do not associate any $t^*$ with those points. The white regions correspond to the points where the spectrum is either purely or partially real, so there is no $t^*$  for these regions. }
    \label{fig:3}
\end{figure}

Given Eq.~\eqref{eq:b5}, we can characterize how $\mathcal{C}(t)$ reaches $(\mathcal{C}_\Omega)_{\text{V}} = 1/2$ with the Krylov spread fidelity \cite{guerra2025correlationskrylovspreadnonhermitian}
\begin{equation}\label{eq:b5.1}
    \mathcal{F}(t) = \abs{\mathcal{C}(t) - \mathcal{C}_\Omega}
\end{equation}
by requiring $ \mathcal{F}(t) < \epsilon < 1$ and then finding the time $t(\epsilon)$ for which the inequality holds. Hence, we denote  
$$\mathcal{C}_{\text{st}}(t)  = \mathcal{C}\big(t>t(\epsilon)\big) \simeq \mathcal{C}_\Omega$$ 
as the stationary-state spread. 
Once $t(\epsilon)$ is found, we find the limit
\begin{equation}
    t^* = \lim_{\epsilon \to 0}\dfrac{t(\epsilon)}{\abs{\ln \epsilon}}
\end{equation}
to obtain an $\epsilon$-independent description, which serves two things: it defines the stationary state for times $t\gg t*$, and it serves as a probe for detecting dynamical phase transitions. By following this description, we find that $t^*$ can attain two values (see Appendix~\ref{sec:krylov_fidelity}):
\begin{widetext}
    \begin{alignat}{3}
        t_1(\epsilon) &\simeq \dfrac{1}{4\sqrt{\gamma^2-4J^2}}W_0 \left( \frac{4(\gamma^2 - 4J^2)^3}{\epsilon^2 \pi \gamma^4(4J^2-h^2)} \right)\ 
    \underbrace{\rightarrow}_{\epsilon\to 0} \ \dfrac{1}{2\sqrt{\gamma^2-4J^2}}\ln \dfrac{1}{\epsilon}, \qquad && t^*_1 = \dfrac{1}{2\sqrt{\gamma^2-4J^2}}, \label{eq:b7.1} \\
    t_2(\epsilon) &\simeq \dfrac{1}{4\sqrt{\gamma^2-h^2}}W_0 \left( \frac{4(\gamma^2 - h^2)^3}{\epsilon^2 \pi \gamma^4(h^2-4J^2)} \right)\ 
    \underbrace{\rightarrow}_{\epsilon\to 0} \ \dfrac{1}{2\sqrt{\gamma^2-h^2}}\ln \dfrac{1}{\epsilon}, \qquad && t^*_2 = \dfrac{1}{2\sqrt{\gamma^2-h^2}}, \label{eq:b7.2}
    \end{alignat}
\end{widetext}
where $W_0(z)$ is the Lambert $W$ function, which has the asymptotics $W_0(z\rightarrow \infty) \simeq \ln z - \ln(\ln z).$ These times refine Region V as shown in Fig.~\ref{fig:3}, where $t_1^*$ corresponds to $\{\abs{h}\leq 2J, \abs{\gamma}>2J\}$ and $t_2^*$ corresponds to $\{\abs{h}\geq 2J, \abs{\gamma} > h\}$.  
Qualitatively, the change in the time $t^*$ from region 1 to 2 is due to the location of the slowest dissipation mode of $\Lambda(k) = i\Gamma(k)$: in Region 1, $\abs{\Gamma(0)}=0$, whereas in Region 2, $\abs{\Gamma(\pi)}=0$. This can be seen in Figs.~\ref{fig:1}b-d in the purple, dotted lines. Due to this change in the slowest dissipation mode as Region V is traversed, the first derivative of $t^*$ with respect to $h$ is discontinuous, indicating thus the presence of the dynamical phase transition.

Let us consider the other regions where the spectrum can have a non-vanishing real part. To address this, we consider the time-average of the spread density
\begin{equation}\label{eq:av1}
   \bar{\mathcal{C}}  = \lim_{T\to \infty}\dfrac{1}{T}\int_{0}^T \dd t \, \mathcal{C}(t), 
\end{equation}
where $\mathcal{C}(t)$ is as in Eq.~\eqref{eq:b5}. The order of the limits in Eq.~\eqref{eq:av1} can be interchanged as the integrand of $\mathcal{C}(t)$ behaves well (see Appendix~\ref{sec:time_average spread density}). Now, in regions III and IV, where the spectrum is complex, one could naively expect $\bar{\mathcal{C}} = \mathcal{C}_\Omega$, where the latter spread is given by Eq.~\eqref{eq:15.1}. However, this does not hold because the continuum of imaginary gapless modes appears whenever $E(k)\neq 0$. The above equality holds in the non-Hermitian Ising model studied in Ref.~\cite{guerra2025correlationskrylovspreadnonhermitian} owing to the presence of only two gapless modes in the imaginary part of the spectrum. Hence, as an educated guess,  $\bar{\mathcal{C}} = \mathcal{C}_\Omega$ holds only when there is a finite number of gapless modes in the imaginary part of the spectrum. This can be seen in Fig.~\ref{fig:2.1} were we show $1-\bar{\mathcal{C}}$. In Region V, where the spectrum is imaginary and can have at most two gapless modes, the two spreads match and equal $1/2$. In the remaining regions, the spreads do not coincide (see Fig.~\ref{fig:2}). However, this averaged spread could also serve as a probe for the phase transitions we explored before with $\mathcal{C}_\Omega$, as the boundaries corresponding to regions II and IV are not smooth. The overall resemblance of $1-\bar{\mathcal{C}}$ and $\mathcal{C}_\Omega$ requires further research.

\section{Summary and discussion}\label{sec:summary}

In this work, we implemented the Krylov spread density as a probe to detect a $\mathcal{PT}$-symmetry breaking in a non-Hermitian SSH model, where the spectrum goes from real to complex, and another spectral transition when the spectrum goes from complex to purely imaginary. In the $\mathcal{PT}$-broken phase, where the spectrum is imaginary, we used the Krylov fidelity to determine how the spread reaches its stationary state limit and found two characteristic times that define two dynamical phases. 

As we saw in Sec.~\ref{sec:spread_complexity}, the spread depends on the initial state and the generator of the dynamics. Therefore, different initial states can yield different spreads for the same Hamiltonian. Hence, since we aimed to study the non-Hermitian vacuum $\ket{\Omega}$ of the non-Hermitian SSH Hamiltonian \eqref{eq:1}, we chose an appropriate initial state $\ket{\psi(0)} = \ket{\text{GS}}$ [Eq.~\eqref{eq:10}] for which $\ket{\Omega} = \Omega \ket{\text{GS}}$ [Eq.~\eqref{eq:12}] is a generalized coherent state. Thanks to this relation, we could find the Hermitian Hamiltonian \eqref{eq:13} inducing the evolution of $\ket{\psi(0)}$ to $\ket{\Omega}$ at $t = 1$. We calculated the spread density $\mathcal{C}_\Omega$ [Eq.~\eqref{eq:15.1}] of this evolution when the non-Hermitian vacuum was reached. The first important conclusion is the lack of symmetry in $\mathcal{C}_\Omega$ with respect to the reflection of the parameters $(h,\gamma) \rightarrow (-h,-\gamma)$ (see Fig.~\ref{fig:2}). In contrast, the spectrum \eqref{eq:7} is symmetric with respect to this reflection (see also Fig.~\ref{fig:1}). Secondly, even though $\mathcal{C}_\Omega$ is continuous across the $h$-$\gamma$ plane, its first and second derivatives with respect to $\gamma$ and $h$ are either discontinuous or diverge when $\mathcal{PT}$-symmetry breaks [Eqs.~\eqref{eq:di1.0}-\eqref{eq:di1}] and when the spectrum becomes purely imaginary [Eqs.~\eqref{eq:di2}-\eqref{eq:di3}]. With this, we can assert the usefulness of the spread as a probe for detecting phase transitions in general. 

The other spread we considered was of the same initial state $\ket{\text{GS}}$ but this time evolving under the non-Hermitian SSH Hamiltonian \eqref{eq:1}. In the $\mathcal{PT}$-broken phase corresponding to an imaginary spectrum, $\ket{\text{GS}}$ evolves to $\ket{\Omega}$ in the infinite-time limit, proving, thus, the importance of that particular initial state as it is related to the vacuum of $H$ in two different ways. Since the same state is reached via the unitary dynamics mentioned before, the time-dependent spread \eqref{eq:b5} of the non-Hermitian dynamics tends to $\mathcal{C}_\Omega$ in the infinite-time limit. By implementing the Krylov fidelity \eqref{eq:b5.1} to study how $\mathcal{C}(t)$ approaches $\mathcal{C}_\Omega$ in this region of the $\mathcal{PT}$-broken phase where the spectrum is imaginary, we were able to determine two times, Eqs.\eqref{eq:b7.1} and \eqref{eq:b7.2}, characterizing two previously unknown dynamical phases existing in this subregion. Interestingly, in this subregion, $\mathcal{C}(t\rightarrow \infty) = \mathcal{C}_\Omega$ is constant, and the entanglement entropy calculated in the $\ket{\Omega}$ same state obeys an area law \cite{legal}. However, having a constant spread does not necessarily imply that the entanglement entropy obeys an area law. In Fig.~\ref{fig:4}, we can see a constant spread for $h<0$ and $\gamma = 0$, where the $\mathcal{PT}$-symmetry is unbroken. In this phase, the entanglement entropy obeys a volume law. 
Furthermore, in Ref.~\cite{guerra2025correlationskrylovspreadnonhermitian}, it was shown that the spread of a non-Hermitian Ising chain is not constant in the region where the entanglement entropy calculated in its vacuum obeys an area law \cite{zerba_measurement_2023}. More research is needed to clarify the relationship between these two quantities.

We calculated the time-averaged spread~\eqref{eq:av1} in the regions where the spectrum is complex. We found that $\bar{\mathcal{C}} \neq \mathcal{C}_\Omega$ except for the region where the spectrum is imaginary. This should be of no surprise, as the non-Hermitian SSH model has the peculiarity of having either a real or imaginary spectrum per each momentum mode $k$. Thus, the absence of those dissipation modes whenever the real part is non-zero impedes the convergence of $\ket{\text{GS}}$ to $\ket{\Omega}$, which translates into a different spread even after performing a time-averaging. Nonetheless, $\bar{\mathcal{C}}$ is qualitatively similar (see Fig.~\ref{fig:spreads}) to  $\mathcal{C}_\Omega$ throughout the $h$-$\gamma$ plane. In particular, it also shows singularities in its derivatives at the critical lines. Therefore, either spread can be used as a probe to detect the phase transitions we described above. However, as the time-averaging produced a quite complicated function in $k$, we were unable to derive an analytic expression for $\bar{\mathcal{C}}$ in the closed form; therefore, $\mathcal{C}_\Omega$ is perhaps more convenient for these purposes. 

The conjecture stated in Ref.~\cite{guerra2025correlationskrylovspreadnonhermitian} that the derivatives of the Krylov spread diverge across any quantum phase transitions was also corroborated for the model here treated. Naturally, as a formal proof is worth pursuing, a tour de-Force can be used to verify this for many other phase transitions, such as, e.g., measurement-induced and $\mathcal{PT}$-transitions. In addition, we studied the Krylov spread only in the left eigenbasis of the non-Hermitian Hamiltonian, and it would be reasonable to study this measure on the left and right basis, i.e., by using a bi-orthogonal approach. Interestingly, in the Hermitian SSH model, the spread is constant in the non-trivial topological phase (see Eq.~\eqref{eq:d1.1} and Ref.~\cite{caputa2}). It is worth researching the relation, if any, between topological phases and the spread in other models and their generalizations~ \cite{nair,Lieu,wong_cheuk,He_2021,nehra,Rottoli_2024,longwen}.  
In addition, the non-Hermitian SSH model with several quenches \cite{gautam_spread_2024} can also be investigated.

Before closing the paper, it is worth mentioning that, while finalizing this manuscript, Ref.~\cite{chakrabarti2025dynamicsmonitoredsshmodel} was posted, where the dynamics of the same non-Hermitian SSH model with periodic and open boundary conditions were analyzed in Krylov space by implementing numerical methods.
The authors were able to detect the $\mathcal{PT}$-transition in the time-dependent Krylov spread calculated via the bi-Lanczos algorithm. However, Ref.~\cite{chakrabarti2025dynamicsmonitoredsshmodel} had to go beyond standard Krylov methods and utilize the quantum Fisher information in Krylov space to detect the complex-to-purely-imaginary spectral transition. In contrast, in our work, all these phase transitions were identified by means of the Krylov spread on an equal footing, based on the analytical solution using a canonical state. Moreover, in addition, we disclosed previously hidden dynamical phase transitions using the same Krylov framework.

\section{Acknowledgments}
 Y.G. was supported by the Deutsche Forschungsgemeinschaft (DFG,
German Research Foundation) grant SH 81/8-1 and by
a National Science Foundation (NSF)–Binational Science Foundation (BSF) grant 2023666.

\appendix
\begin{widetext}
\section{Diagonalization of $H$ and the non-Hermitian Bogoliubov vacuum}\label{sec:diag}
In this appendix, we diagonalize the non-Hermitian Hamiltonian~\eqref{eq:1} by implementing the Fourier representation~\eqref{eq:2} and find the eigenvalues~\eqref{eq:7} as well as the ground state~\eqref{eq:12}. We also demonstrate that this ground state is a generalized coherent state of the ground state~\eqref{eq:11}. 

After using the Fourier representation~\eqref{eq:2}, the Hamiltonian can be written as  $H = \sum_k H(k)$, where
\begin{equation}\label{eq:ap1}
    H(k) = \begin{pmatrix}
        c_k^\dagger & d_k^\dagger 
    \end{pmatrix}\begin{pmatrix}
        t_2 \sin k & t_1 +t_2 \cos k + t_2 -\gamma \\
        t_1 + t_2 \cos k +\gamma & -t_2 \sin k
    \end{pmatrix}\begin{pmatrix}
        c_k \\ d_k
    \end{pmatrix}.
\end{equation}
Alternatively, we can set $H(k) \equiv \Psi_k^\dagger \vec R(k)\cdot \vec \sigma\, \Psi_k$, where $\Psi_k = (c_k, d_k)^T$, 
\begin{equation}\label{eq:ap1.1}
    \vec R(k) = \begin{pmatrix}
        R_x \\ R_y \\ R_z
    \end{pmatrix} = \begin{pmatrix}
        t_1+t_2\cos k \\
          -i \gamma \\
        t_2 \sin k
    \end{pmatrix}
\end{equation}
is a complex Bloch vector, and $\vec \sigma =(\sigma_x,\sigma_y,\sigma_z)$ is a vector of Pauli matrices. Recall that $$t_1 = -J-\frac{h}{2}, \, \, t_2 = -J + \frac{h}{2}.$$ The matrix $\vec R(k)\cdot \vec{\sigma}$ can be diagonalized by a similarity transformation as $V_k^{-1}\vec R(k) \cdot \vec{\sigma}\, V_k = \text{diag}\left[\Lambda(k), -\Lambda(k) \right]$, where
\begin{equation}\label{eq:ap2}
    V_k = \begin{pmatrix}
        u_k & u_k \\ v_k^+& v_k^-
    \end{pmatrix} \equiv  \begin{pmatrix}
        \vec{v}_+(k) & \vec{v}_-(k)
    \end{pmatrix},
\end{equation}
and
\begin{equation}\label{eq:ap3}
    \vec{v}_{\pm}(k) = \dfrac{1}{\sqrt{2R(R \mp R_z)}}\begin{pmatrix}
        R_x - i R_y \\ \pm R- R_z
    \end{pmatrix}.
\end{equation}
Here, $R = \Lambda(k)$ is the positive root given by Eq.~\eqref{eq:7}.

Thanks to the above similarity transformation, we can define the non-Hermitian fermionic operators
\begin{equation}\label{eq:ap4}
    X_k^* = \begin{pmatrix}
        \chi_{+,k}^* & \chi_{-,k}^*
    \end{pmatrix} \coloneqq \Psi_k^\dagger V_k = \begin{pmatrix}
        u_k c_k^\dagger + v_k^+d_k^\dagger & u_k c_k^\dagger + v_k^-d_k^\dagger
    \end{pmatrix}
\end{equation}
and
\begin{equation}\label{eq:ap5}
    X_k = \begin{pmatrix}
        \chi_{+,k} \\ \chi_{-,k}
    \end{pmatrix} \coloneqq V_k^{-1} \Psi_k = \dfrac{1}{\det V_k} \begin{pmatrix}
        v_k^- c_k - u_k d_k \\ -v_k^+ c_k + u_kd_k.
    \end{pmatrix},
\end{equation}
which satisfy the anticommutation relations \eqref{eq:5}. Given these new fermionic operators, we can write the Hamiltonian as in Eq.~\eqref{eq:4}

Akin to the Hermitian version of Eq.~\eqref{eq:1}, the unnormalized ground state of $H$ (cf. Eq.~\eqref{eq:12}), which is a non-Hermitian Bogoliubov vacuum, is obtained by populating the vacuum $\ket{0}$, which is annihilated by $c_k$ and $d_k$, by the non-Hermitian quasiparticles associated to a negative complex spectrum, i.e., 
\begin{equation}\label{eq:ap6}
    \ket*{\tilde \Omega} = \prod_{k\in \mathcal{K}^+}\chi_{-,-k}^*\chi_{-,k}^*\ket{0},
\end{equation}
where $\mathcal{K}^+ = \mathcal{K}\cap \mathbb{R}_{\geq 0}.$
 Thus, $H\ket*{\tilde \Omega} = E_{\Omega} \ket*{\tilde \Omega}$, and so the ground state energy is
\begin{equation}\label{eq:ap7}
    E_\Omega = -\sum_{k \in \mathcal K^+}\Lambda(k).
\end{equation}
This is a natural choice of the sign of the spectrum for each $k$, for whenever $\mathfrak{Im}\,\Lambda(k) = \Gamma(k)  \neq 0$, $\ket*{\tilde \Omega}$ decays the slowest. Moreover, if $\Gamma(k)=0$, then $\Lambda(k) = E(k) <0$ is a real number, and Eq.~\eqref{eq:ap6} is the state with the lowest energy in the given non-Hermitian system. 

To demonstrate that $\ket{\Omega} = \norm*{\ket*{\tilde\Omega}}^{-1/2} \ket*{\tilde \Omega}$ a generalized coherent state of Eq.~\eqref{eq:10}, which is the ground state of Eq.~\eqref{eq:11}, let us consider the non-Hermitian Hamiltonian once more, this time, written in a slightly different way:
    \begin{align}\label{eq:ap8}
        H &=\sum_{k \in \mathcal{K}^+}\left[R_z(k)\left(c_k^\dagger c_k - d_k^\dagger d_k \right) + R_+ c_k^\dagger d_k + R_- d_k^\dagger c_k-R_z(k)\left(c_{-k}^\dagger c_{-k} - d_{-k}^\dagger d_{-k} \right) + R_+ c_{-k}^\dagger d_{-k} + R_- d_{-k}^\dagger c_{-k} \right],
    \end{align}
where we used $R_z(-k) = -R_z(k)$ and set $R_+(k) = R_x(k) - iR_y(k)$ and $R_-(k) = R_x(k)+iR_y(k)$. Contrary to the Hermitian case, $R_+^* \neq R_-$. In this form, $H$ can be seen as an element of the semisimple Lie algebra  $\mathfrak{g} = \mathfrak{g}^+\times \mathfrak{g}^- = \mathfrak{su}(2)\times \mathfrak{su}(2)$. More precisely, if we set
\begin{equation}\label{eq:ap9}
    J^+_+(k) \coloneqq c_k^\dagger d_k, \quad J_-^+(k) \coloneqq d_k^\dagger c_k, \quad J_z^+(k) \coloneqq \dfrac{1}{2}\left(c_k^\dagger c_k - d_k^\dagger d_k \right),
\end{equation}
and
\begin{equation}\label{eq:ap10}
    J^-_+(k) \coloneqq d_{-k}^\dagger c_{-k}, \quad J_-^-(k) \coloneqq c_{-k}^\dagger d_{-k}, \quad J_z^-(k) \coloneqq -\dfrac{1}{2}\left(c_{-k}^\dagger c_{-k} - d_{-k}^\dagger d_{-k} \right), 
\end{equation}
it holds that
\begin{equation}\label{eq:ap11}
    [J_{+}^s(k), J_{-}^{s'}(k')] =2\delta _{ss'}\delta_{k,k'}J_z^s(k), \quad [J_z^{s}(k), J_\pm^{s'}(k')] = \pm \delta_{ss'}\delta_{kk'}J_\pm^s(k), 
\end{equation}
where $s,s'=\pm$. The superindices denote the Lie algebra to which the element belongs, and the subindices denote the generator in the given Lie algebra. Thus, Eq.~\eqref{eq:ap8} can be rewritten as 
\begin{equation}\label{eq:ap12}
    H = \sum_{\substack{k \in \mathcal{K}^+ \\ s=\pm}} H^{(s)}(k),
\end{equation}
where $$H^{(s)}(k) = 2R_z(k) J_z^s(k) + R_+^s(k)J_+^s(k)+R_-^s(k)J_-(k).$$ Note that $R^+_+ = R_-^- = R_+$ and $R^+_- = R_+^-= R_-$. 

Given this Lie-algebraic language, and by noting that $J_{-}^{\pm}(k)\ket{\text{GS}} = 0$, we can identify $\ket{\text{GS}}$ with the lowest-weight state of $\mathfrak{g}$, 
\begin{equation}\label{eq:ap13}
    \ket{\text{GS}} = \prod_{k\in \mathcal{K}^+}c_{-k}^\dagger d_k^\dagger\ket{0}=\bigotimes_{k \in \mathcal{K}}\ket{1/2,-1/2}_{k},
\end{equation}
for which we have
$$J_{-}^{\pm}(k)\ket{1/2,-1/2}_{\pm k }=0,\quad J_+^{\pm}(k)\ket{1/2,-1/2}_{\pm k}=\ket{1/2,1/2}_{\pm k}, \ \ \text{and} \ \ J_z^{\pm}(k)\ket{1/2,\pm 1/2}_{\pm k} = \pm \frac{1}{2}\ket{1/2,\pm 1/2}_{\pm k}.$$ Moving forward, by noticing that $[J^{\pm }_+(k), c_{-k'}^\dagger d_{k'}^\dagger]= 0$ for $k\neq k'$, we can analyze the action of $J_{\pm}^\pm (k)$ on $\ket{\text{GS}}$ by focusing on the following states:
\begin{align}
    J_+^+(k)c_{-k}^\dagger d_k^\dagger\ket{0} &=  c_{-k}^\dagger c_k^\dagger \ket{0},\label{eq:ap14} \\
    J_+^-(k)c_{-k}^\dagger d_k^\dagger\ket{0} &= d_{-k}^\dagger d_k^\dagger \ket{0},\label{eq:ap15}
\end{align}
and
\begin{equation}\label{eq:ap16}
        J_+^+(k)J_+^-(k)\ket{0} = d_{-k}^\dagger c_k^\dagger \ket{0}.
\end{equation}
Now, for $y_\pm(k)\in \mathbb{C}$ and $k\in \mathcal{K}^+$, let
\begin{equation}
\tilde \Omega_k \coloneqq \exp[y_+(k)J^+_+(k)+y_-(k)J_+^-(k)]
\end{equation}
and consider
\begin{equation}
     \tilde \Omega_k \,  c_{-k}^\dagger d_k^\dagger\ket{0} = \exp[y_+(k)J_+^+(k)+y_-(k)J_+^-(k)]c_{-k}^\dagger d_k^\dagger\ket{0} =\left(c_{-k}^\dagger d_k^\dagger +  y_+c_{-k}^\dagger c_k^\dagger + y_-d_{-k}^\dagger d_k^\dagger + y_+ y_-d_{-k}^\dagger c_k^\dagger  \right)\ket{0}.
\end{equation}
On the other hand, by expanding Eq.~\eqref{eq:ap6}, we get
\begin{equation}\label{eq:ap17}
     \ket*{\tilde \Omega} = \prod_{k\in \mathcal{K}^+}u_{-k}v_k^{-}\left(c_{-k}^\dagger d_k^\dagger + \dfrac{u_k}{v_k^-} c_{-k}^\dagger c_k^\dagger + \dfrac{v_{-k}^-}{u_{-k}} d^\dagger_{-k} d_k^\dagger + \dfrac{u_k v_{-k}^-}{u_{-k}v_k^-}d_{-k}^\dagger c_k^\dagger  \right)\ket{0}.
\end{equation}
Hence, if we set 
\begin{equation}\label{eq:ap18}
y_+(k) =  \dfrac{u_k}{v_k^-} \quad \text{and} \quad y_-(k) =  \dfrac{v_{-k}^-}{u_{-k}}    
\end{equation}
we can write Eq.~\eqref{eq:ap17} as
\begin{equation}\label{eq:ap19}
    \ket*{\tilde \Omega} = \prod_{k\in \mathcal{K}^+}u_{-k}v_k^{-}\exp(\dfrac{u_k}{v_k^-}J_+^+(k) +\dfrac{v_{-k}^-}{u_{-k}} J_+^-(k) )\ket{\text{GS}},
\end{equation}
and, after normalizing, it reads 
\begin{equation}\label{eq:ap20}
    \ket{\Omega} = \prod_{k\in \mathcal{K}^+}\dfrac{u_{-k}v_k^{-}}{\abs{u_{-k}v_k^{-}}} \dfrac{\exp[y_+(k)J^+_+(k)+y_-(k)J_+^-(k)]}{\sqrt{\left(1 + \abs{y_+(k)}^2\right)\left(1 + \abs{y_-(k)}^2\right)}
    }
 \ket{\text{GS}}.
\end{equation}

As our final step, let $\alpha_z, \alpha_\pm \in \mathbb{C}$, and $J_\pm, J_z \in \mathfrak{su}(2)$, and consider the normal-ordering $\mathfrak{su}(2)$ decomposition formula \cite{ban1993decomposition}
\begin{equation}\label{eq:ap23}
    \exp(a_+J_+ + a_zJ_z + a_- J_-) = \exp(\alpha_+ J_+)\exp[\log(\alpha_z)J_z]\exp(\alpha_- J_-),
\end{equation}
with
\begin{equation}
    \alpha_{\pm} = \dfrac{a_{\pm }}{\phi}\dfrac{\sinh \phi}{\cosh \phi - \dfrac{a_z}{2\phi}\sinh \phi}, \quad
    \alpha_z = \left[\cosh \phi - \dfrac{a_z}{2\phi}\sinh \phi \right]^{-2}, \quad
    \phi = \sqrt{\dfrac{a_z^2}{4}+a_+a_-},
\end{equation}
applied to the following unitary operators
\begin{equation}\label{eq:ap24}
    \Omega_k^+ \coloneqq \exp[y_+(k)J_+^+(k)- \text{H.c}], \quad \text{and} \quad 
    \Omega_k^- \coloneqq \exp[y_-(k)J_+^-(k)- \text{H.c}]
\end{equation}
acting on $\ket{1/2,-1/2}_{k}$ and $\ket{1/2,-1/2}_{-k}$, respectively. Upon applying those steps and some simple algebraic manipulations, we get
\begin{align}
    \Omega_k^\pm \ket{1/2,-1/2}_{\pm k} &= \dfrac{\exp[y_\pm(k) J_+^\pm(k)]}{\sqrt{1+\abs{y_\pm(k)}^2}}\ket{1/2,-1/2}_{\pm k}.
\end{align}
Thus, by calling $\Omega \coloneqq \Omega_k^+\Omega_k^-$ and $\exp[i\varphi(\Omega)] \coloneqq \prod_{k \in \mathcal{K}^+}u_{-k}v_k^{-}\abs{u_{-k}v_k^{-}}^{-1}$, we can immediately conclude that Eq.~\eqref{eq:ap20} is equal to 
\begin{equation}\label{eq:ap26}
    \ket{\Omega} = \exp[i\varphi(\Omega)]\Omega \ket{\text{GS}}.
\end{equation}
Finally, noting that the action of operators of the form $\exp[a_z^+J_z^+(k)+a_z^-J_z^-(k)]$ on $\ket{\text{GS}}$ only produces a phase, we can conclude that $\ket{\Omega}$ is a generalized coherent state of $\ket{\text{GS}}$ defined in the coset space $\text{SU}(2)^{\otimes L}/\text{U}(1)^{\otimes L}$ \cite{perelomov_generalized_1986,kam_coherent_2023}, and thus the phase appearing in Eq.~\eqref{eq:ap26} is not important in our analysis. Note also that the Hermitian Hamiltonian~\eqref{eq:13} used in the unitary evolution $\ket{\psi(1)} = e^{-iH_\Omega}\ket{\text{GS}}$ is obtained from $\Omega$ [see Eq.~\eqref{eq:ap24}]. Indeed, for $s = \pm,$ consider $\Omega_k^s = \exp[-i\left(i y_s(k) J_+^s(k) + \text{H.c.} \right)]$. Then, $$\Omega = \Omega_k^+\Omega_k^- = \exp[-i\sum_{s=\pm}\Big(i y_s(k) J_+^s(k) + \text{H.c.}  \Big)] = \exp(-iH_\Omega),$$ with $H_\Omega$ given by Eq.~\eqref{eq:13}.

\section{Time-evolution of $\ket{\text{GS}}$  under the non-Hermitian SSH Hamiltonian}\label{sec:time_evolution_GS}
 In what follows, we find the time-evolution of the state $\ket{\text{GS}}$ induced by $H$ by invoking the same $\mathfrak{su}(2)$ decomposition formula \eqref{eq:ap23} and obtain Eq.~\eqref{eq:b2}. To this end, let $s = \pm$ and consider 
\begin{subequations}
\begin{align}\label{eq:app29.0}
    \exp[-iH^s(k)t]&= \exp{-i\left[2R_z(k)J_z^s(k) + R_+^s(k)J_+^s(k) + R_-^s(k)J_-^s(k) \right]t}\\
    &= \exp[A_+^s(k;t)J_+^s(k)]\exp[\ln (A_z^s(k;t))J_z^s(k)]  \exp[A_-^s(k;t)J_-^s(k)],
\end{align}
\end{subequations}
where
\begin{align}\label{eq:app29}
    A_{\pm}^s(k;t) &= -i\dfrac{R_s^s(k)}{\Lambda(k)} \dfrac{\sin \Lambda(k)t}{\cos \Lambda(k)t + i[R_z(k)/\Lambda(k)]\sin \Lambda(k)t}
\end{align}
and
\begin{equation}\label{eq:app30}
    A_{z}^+(k;t) = \Bigl( \cos \Lambda(k)t + i[R_z(k)/\Lambda(k)]\sin \Lambda(k)t \Bigr)^{-2}.
\end{equation}
Thus, up to a phase, the time-evolution of $\text{GS}$ reads
\begin{equation}\label{eq:app33}
    \ket*{\psi(t)}  =  \dfrac{e^{-iHt}\ket{\text{GS}}}{\norm{e^{-iHt}\ket{\text{GS}}}}  =   \prod_{\substack{k>0 \\ s=\pm} } \dfrac{\exp[A_+^{s}(k;t)J_+(k)]}{\sqrt{1+\abs{A_+^s(k;t)}^2}}\ket{\text{GS}}.
\end{equation}
Given this expression, let us assume that $\Lambda(k) = i\Gamma(k) \in i\mathbb{R}$, where $\Gamma(k) = \sqrt{\abs{R_y(k)}^2-R_x^2(k)-R_z^2(k)}$. Thus, we can conveniently rewrite $A_+^{\pm }(k;t)$ as
\begin{equation}
    A_+^{\pm }(k;t) = -\dfrac{R_x(k) \mp i R_y(k)}{R_z(k)+ i\Gamma(k)}\dfrac{1}{1-l_k(t)},
\end{equation}
where
\begin{equation}
    l_k(t) \coloneqq \frac{2}{f_k\left( 1 - \exp[-2\Gamma(k)t] \right)}
\end{equation}
and $f_k \coloneqq 1+R_z(k)[i \Gamma(k)]^{-1}$. In the limit $t \rightarrow \infty$, $l_k(t)\rightarrow 2f_k^{-1}$, and $A_+^{\pm}(k;t) \rightarrow x_\pm(k)$ (see Eq.~\eqref{eq:b3}). Thus, Eq.~\eqref{eq:app33} yields Eq.~\eqref{eq:b2.1} in the limit $t \rightarrow \infty$.

Now, from Eq.~\eqref{eq:ap18}, let us note that 
\begin{equation}
    y_+(k) = \dfrac{u_k}{v_k^-} = \dfrac{R_x(k) - i R_y(k)}{R_z(k) + i\Gamma(k)} \quad \text{and} \quad    y_-(k) = \dfrac{v_{-k}^-}{u_{-k}} = \dfrac{R_x(k) + i R_y(k)}{R_z(k) + i \Gamma(k)}.
\end{equation}
Hence, $y_\pm(k)$ and $x_\pm(k)$ differ by phase, which implies that $\ket{\psi(t\rightarrow \infty)}$ can be related to $\ket{\Omega}$ via the unitary operator \eqref{eq:b4}. 

\section{Krylov fidelity}\label{sec:krylov_fidelity}
In this section, we outline the steps that lead to the times \eqref{eq:b7.1} and \eqref{eq:b7.2} obtained from $\mathcal{F}(t) <\epsilon$. 

By expanding Eq.~\eqref{eq:b5.1}, we get
\begin{align}\label{eq:o1}
    \mathcal{F}(t) & = \abs{\int_{0}^\pi\dfrac{\dd k}{2\pi}\left( \dfrac{\abs{A_+^+(k;t)}^2}{1+\abs{A_+^+(k;t)}^2} - \dfrac{1}{1+\abs{A_+^-(k;t)}^2} \right)}.
\end{align}
Upon a few algebraic manipulations, we get
$$\abs{A_+^\pm(k;t)}^2 = \dfrac{R_z^2(k) + \Gamma^2(k)}{R_z^2(k) + \Gamma^2(k) \coth^2 \Gamma(k) t},$$
and, given this expression, for the points belonging to the set $\{\abs{\gamma}>2J, 0<\abs{h} < 2J \}$, $\abs{\Gamma(k)}$ has a minimum at $k=0$, which corresponds to the slowest decay mode, and the integrand of Eq.~\eqref{eq:o1} is concentrated around this point. Hence, for sufficiently large $\Gamma(k) t$
\begin{align}
    \mathcal{F}(t) &\approx  \abs{\dfrac{\gamma^2-4J^2}{\pi \gamma^2}e^{-2\sqrt{\gamma^2-4J^2}t} \int_0^\infty \dd k \, \exp[ -\dfrac{4J^2-h^2}{4\sqrt{\gamma^2-4J^2}} k^2 t  ]} \notag \\
    &= \dfrac{(\gamma^2-4J^2)^{5/4}}{\sqrt{4J^2-h^2}\, \gamma^2}\dfrac{e^{-2\sqrt{\gamma^2-4J^2}\,t}}{\sqrt{\pi t}}.
\end{align}
For a given $\epsilon \in (0,1)$, we invert $\mathcal{F}(t) < \epsilon$ and get Eq.~\eqref{eq:b7.1}. 

Turning to the region $\{\abs{\gamma}>\abs{h}, \abs{h}>2J \}$, the minimum of $\abs{\Gamma(k)}$ shifts to $k=\pi$ and the integrand of Eq.~\eqref{eq:o1} concentrates at this point. Thus, under similar steps,
\begin{align}
    \mathcal{F}(t) &\approx \abs{ \dfrac{h^2-\gamma^2}{\pi \gamma^2}e^{-2\sqrt{\gamma^2-h^2}t}\int_{-\infty}^\pi \dd k \, \exp[\dfrac{-(h^2-4J^2)}{4\sqrt{\gamma^2-h^2}}(k-\pi)^2t]}\notag \\
    &= \dfrac{(\gamma^2-h^2)^{5/4}}{\sqrt{h^2-4J^2}\, \gamma^2}\dfrac{e^{-2\sqrt{\gamma^2-h^2}\,t }}{\sqrt{\pi  t}}.
\end{align}
Once again, by fixing an $\epsilon \in (0,1)$ and setting $\mathcal{F}(t) <\epsilon$, one can invert the resulting expression to find the time \eqref{eq:b7.2}.

\section{Time-averaged spread density}\label{sec:time_average spread density}
In this section, we demonstrate that the order of the integrals in Eq.~\eqref{eq:av1} can be interchanged. Namely,
\begin{equation}\label{eq:x1}
    \lim_{T\to \infty}\dfrac{1}{T}\int_0^T\dd t \, \sum_{s=\pm } \int_0^\pi \dfrac{\dd k }{2\pi}  C_s(k;t) =\sum_{s=\pm }  \int_0^{\pi}\dfrac{\dd k }{2\pi} \,\lim_{T\to \infty}\int_0^T \dd t\, C_s(k;t).
\end{equation}
Let $H$ have a purely real spectrum, i.e., $\Lambda(k) = E(k)$, which is also gapped. Then, 
$$C_s(k;t) = \dfrac{\abs{A_+^s(k;t)}^2}{1+\abs{A_+^s(k;t)}^2} = \dfrac{\left(R_x(k) - s \abs{R_y(k)} \right)^2 \sin^2E(k)t}{\left[ \left(R_x(k) - s \abs{R_y(k)} \right)^2 + R_z^2(k)\right]\sin^2 E(k)t + E^2(k)\cos^2E(k)t}.$$
Let $T_n = 2\pi n/E(k)$ for $n \in \mathbb{Z}^+$ and define the sequence of functions $\{f_n(k) \}_n$ with
$$ f_n(k) = \frac{1}{T_n}\int_{0}^{T_n}\dd t\,\left[ C_1(k;t)+C_2(k;t)\right]$$
converging to 
$$ f(k) =  \lim_{n\to \infty}f_n(k) = \sum_{s=\pm} \dfrac{\left(R_x(k) - s \abs{R_y(k)} \right)^2}{\left(R_x(k) - s \abs{R_y(k)} \right)^2 + R_z^2(k) + \left[ \left(R_x(k) - s \abs{R_y(k)} \right)^2 + R_z^2(k)\left[ R^2_x(k)- R_y(k)^2 + R_z^2(k) \right] \right]^{1/2} }$$
for every $k\in[0,\pi]$. Now, since
$$  \abs{f_n(k)} \leq \dfrac{1}{T_n}\int_0^{T_n} \dd t \abs{C_1(k;t)+C_2(k;t)} \leq \dfrac{1}{T_n}\int_0^{T_n}\dd t\, 2 = 2 \eqqcolon g(k)$$ and 
$$\int_0^\pi \dfrac{\dd k }{2\pi}g(k) = 1$$
for every $n \in \mathbb{Z}^+$, we can use the Dominated Convergence Theorem \cite{axler} and get
$$ \lim_{n\rightarrow \infty }\int_0^\pi \dfrac{\dd k }{2\pi}f_n(k) = \int_{0}^\pi \dfrac{\dd k }{2\pi} \, f(k).$$
If $E(k)$ is gapless at some $k = q$,  $C_s(q;t) = 0$ and so $f_n(q) = 0$ and $f(q) = 0$. 

Let $H$ now have a purely imaginary spectrum, i.e., $\Lambda(k) = i \Gamma(k)$, that is also gapped. Then, 
$$C_s(k;t) = \dfrac{\left(R_x(k) - s \abs{R_y(k)} \right)^2 \sinh^2\Gamma(k)t}{\left[ \left(R_x(k) - s \abs{R_y(k)} \right)^2 + R_z^2(k)\right]\sinh^2 \Gamma(k)t + \Gamma^2(k)\cosh^2\Gamma(k)t}.$$
 Under similar steps as before, but this time with $f(k) = 1,$ we have 
 $$ \lim_{n\rightarrow \infty }\int_0^\pi \dfrac{\dd k }{2\pi}f_n(k) = \int_{0}^\pi \dfrac{\dd k }{2\pi} \, f(k) = \dfrac{1}{2}.$$

Now, for a complex spectrum, we only need to separate the region of integrations accordingly and apply the same analysis as above.  Finally, by noting the equivalence 
$$ \lim_{T\rightarrow \infty}\dfrac{1}{T}\int_0^T\dd t \left[ C_1(k;t) + C_2(k;t) \right] = \lim_{n\rightarrow \infty}  \dfrac{1}{T_n}\int_0^{T_n}\dd t \left[ C_1(k;t) + C_2(k;t) \right],$$
we have that Eq.~\eqref{eq:x1} holds.

\end{widetext}

\bibliography{articles.bib}

\end{document}